\documentclass[iop]{emulateapj}
\usepackage{natbib}
\usepackage{amsmath}
\usepackage{textcomp}
\usepackage[colorlinks=true,citecolor=blue]{hyperref}
\usepackage{graphicx,color}
\usepackage{times}
\usepackage{amssymb}

\newcommand{\lum}{erg\,s$^{-1}$}
\newcommand{\phflux}{\mbox{${\rm \, ph \,\, cm^{-2} \, s^{-1}}$}}
\newcommand{\ergflux}{\mbox{${\rm \, erg \,\, cm^{-2} \, s^{-1}}$}}

\slugcomment{ApJ Letters accepted}

\shorttitle{The brightest GeV Flare of 3C 279}
\shortauthors{Vaidehi S. Paliya}

\begin{document}

\title{{\it Fermi}-Large Area Telescope Observations of the Exceptional Gamma-Ray Flare from 3C 279 in 2015 June}

\author{Vaidehi S. Paliya} 
\affil{Indian Institute of Astrophysics, Block II, Koramangala, Bangalore-560034, India}
\affil{Department of Physics, University of Calicut, Malappuram-673635, India}
\email{vaidehi@iiap.res.in}

\begin{abstract}
An exceptional $\gamma$-ray outburst from 3C 279 is detected by {\it Fermi}-Large Area Telescope (LAT) in 2015 June. In the energy range of 0.1$-$300 GeV, the highest flux measured is (39.1$\pm$2.5) $\times$ 10$^{-6}$ \phflux, which is the highest $\gamma$-ray flux ever detected from 3C 279, exceeding the previous historically brightest flare observed by {\it EGRET} in 1996. The high activity period consists of three major flares with the last one being the brightest. All but one flares show a faster rise and slower decay pattern and at the peak of the activity, the $\gamma$-ray spectrum is found to show a clear signature of break/curvature. The obtained spectral parameters hint for the peak of the inverse Compton emission to lie in the LAT energy range (around $\sim$1 GeV) which is in contrast to that seen during the 2013 December and 2014 April $\gamma$-ray flares of 3C 279. From the $\gamma\gamma$ pair opacity arguments, the minimum Doppler factor is estimated to be 14 and the location of the $\gamma$-ray emitting region is found to be either at the outer edge of the broad line region or farther out from it.

\end{abstract}


\section{Introduction}\label{sec:intro}
The radio source 3C 279 \citep[$z$=0.536;][]{1965ApJ...142.1667L} is one of the best studied flat spectrum radio quasars (FSRQ) in all the wavebands. It emits strong and variable emission at all frequencies and is one of the first blazars detected by {\it EGRET} instrument on the {\it Compton Gamma-Ray Observatory} \citep{1992ApJ...385L...1H}. Since 2008, 3C 279 is being continuously monitored by {\it Fermi}-Large Area Telescope (LAT) and was a subject of various multi-wavelength campaigns \citep[e.g.][]{2012ApJ...754..114H,2015arXiv150204699H}. This source is an active $\gamma$-ray emitter and multiple episodes of $\gamma$-ray outbursts have been detected \citep[][]{2012ApJ...754..114H,2015ApJ...803...15P}. In particular, two prominent flares of similar intensity ($F_{\rm GeV}\sim12$; where $F_{\rm GeV}$ is in 0.1$-$300 GeV range and in units of 10$^{-6}$ \phflux) have been detected by LAT in 2013 December and 2014 April. Interestingly, the 2013 December flare exhibited a hard rising spectrum, whereas a significant curvature was noticed in 2014 April flare \citep{2015arXiv150204699H,2015ApJ...803...15P}. A $\gamma$-ray flux variability as short as $\sim$1 hr is also reported \citep{2015ApJ...803...15P}.

In 2015 June, 3C 279 displayed an exceptionally high activity state. The daily binned $\gamma$-ray flux is reported as the highest measured since the beginning of the {\it Fermi} operation \citep{2015ATel.7633....1C}. Two target of opportunity observations were approved for a total of 750 ksec duration (between MJD 57,188 and 57,197) and during this period {\it Fermi} observed 3C 279 in a special pointed mode, other than its normal sky scanning mode operation. In this letter, motivated by the availability of good quality LAT data, the intraday $\gamma$-ray variability and associated spectral changes are studied in detail. The obtained results are further compared with the previous major flares detected from 3C 279. In Section~\ref{sec:fermi}, observations and data analysis of {\it Fermi}-LAT data from 2015 June 7 to June 20 (MJD 57,180$-$57,193) are reported. Results are presented in Section~\ref{sec:results} and discussed in Section~\ref{sec:dscsn}. Throughout, a $\Lambda$CDM cosmology with the Hubble constant $H_0=71$~km~s$^{-1}$~Mpc$^{-1}$, $\Omega_m = 0.27$, and $\Omega_\Lambda = 0.73$ is adopted.

\section{{\it Fermi}-Large Area Telescope Observations}\label{sec:fermi}
To analyze the LAT data (P7REP) covering the period of $\gamma$-ray outburst, the standard data reduction procedure is adopted and here it is described in brief. The unbinned likelihood method included in the pylikelihood library of {\tt Science Tools (v9r33p0)} is used to extract the information from the events belonging to the SOURCE class and energy range 0.1$-$300 GeV and lying within 10$^{\circ}$ region of interest (ROI) centered at the position of 3C 279. A maximum likelihood (ML) test statistic TS=2$\Delta$log($\mathcal{L}$), where $\mathcal{L}$ represents the likelihood function between models with and without a point source at the position of source of interest, is calculated to determine the significance of the $\gamma$-ray signal. All the sources lying within the ROI and defined in the third {\it Fermi}-LAT catalog \citep[3FGL;][]{2015ApJS..218...23A} are included and their spectral parameters are left free to vary during the model fitting. Sources lying within 10$^{\circ}$ to 15$^{\circ}$ from the center of the ROI are also appeared in the model file and their parameters are kept fixed to the 3FGL catalog values. A first run of the ML analysis is performed over the period of interest and all the sources with TS$<$25 are removed from further analysis.

To characterize the variability properties of the source, light curves are generated using various time binnings (3 hr, 6 hr, 12 hr, and 1 day) and over different energy intervals (0.1$-$300 GeV, 0.1$-$1 GeV, and 1$-$300 GeV). Though the $\gamma$-ray spectral shape of 3C 279 exhibit a definite curvature, all the light curves are produced by adopting a simple power law (PL) in each time bin, since the statistical uncertainties on the PL indices are smaller compared to those obtained from complex model fits such as broken power law (BPL).

Various models are applied to fit the $\gamma$-ray spectra of 3C 279 and that includes a BPL ($N(E) = N_0(E/E_{\rm break})^{-\Gamma_i},$ with $i = 1$ if $E < E_{\rm break}$ and $i = 2$ if $E > E_{\rm break}$), a log-parabola ($N(E) = N_0(E/E_p)^{-\alpha-\beta~{\rm log}(E/E_p)}$, where $\alpha$ is the photon index at $E_p$, $\beta$ is the curvature index and $E_p$ is fixed at 300 MeV), and a PL model over logarithmically equally spaced energy bins with $\Gamma$ kept fixed to the value fitted over the whole energy range. 

All the time/energy bins with $\Delta F_{\gamma}/F_{\gamma} > 0.5$, where $\Delta F_{\gamma}$ is the error estimate in the flux $F_{\gamma}$, and/or TS$<$9\footnote{TS of 9 corresponds to $\sim$3$\sigma$ detection \citep{1996ApJ...461..396M}.} are rejected from the analysis. Systematics on the measured fluxes are of around 10\% below 100 MeV, 5\% between 316 MeV to 10 GeV and 10\% above 10 GeV. The statistical uncertainties are estimated at 1$\sigma$ level.

\section{Results}\label{sec:results}
The daily binned light curve of 3C 279 at photon energies 0.1$-$300 GeV, covering the period of high activity is shown in the top panel of Figure~\ref{fig:fermi_lc}. The source started showing activity around MJD 57,184 with flux level about $F_{\rm GeV}\sim$1 and after a period of 3 days, $\gamma$-ray flux increased by a factor of 10 where it remain steady for 2 days. The source displayed another flare of higher amplitude on MJD  57,189 when the highest $\gamma$-ray flux obtained as $F_{\rm GeV}$ =  24.5$\pm$0.5, which is the highest daily binned $\gamma$-ray flux measurement from 3C 279 since the launch of {\it Fermi}-LAT. The associated photon index is hard and having a value of 2.05$\pm$0.02. Immediately after this flaring activity, the source returned to a relatively low activity state where flux remained at the level of $F_{\rm GeV}\sim$1. Based on the observed activity in the daily binned light curve, the entire period is then divided into four states: a pre$-$flare, two flares, and a post$-$flare state. These periods are shown in the top panel of Figure~\ref{fig:fermi_lc} and labeled as Pre, P1, P2, and Post, respectively.

Due to good photon statistics, the light curves are also generated using time bins of 12 hr, 6 hr, and 3 hr and are shown in the bottom three panels of Figure~\ref{fig:fermi_lc}. As can be seen, the flare of period P1 starts getting resolved in finer bin light curves. In fact, this period comprises of two flares distinctly visible in 6 hr binned light curve. There is a third sharp flare as P1 ends and P2 begins. These three flares are named as F1, F2, and F3 for the rest of the analysis (see Figure~\ref{fig:fermi_lc}). The rising phase of the brightest flare F3 is clearly unresolved down to 3 hr scale whereas its decaying segment seems to be nicely resolved in 6 hr binning. Though the flares F1 and F2 appear to be resolved in 6 hr binning, the 3 hr binning light curve suggests for the presence of sub-structures consisting of several large amplitude yet unresolved and exhibiting chaotic events. The highest $\gamma$-ray flux, using 3 hr binning, is obtained as $F_{\rm GeV}$ = 33.5$\pm$1.4. Moreover, the data is also analyzed using the time bins defined as Good Time Intervals \citep[GTI,][]{2011A&A...530A..77F}. The highest $\gamma$-ray flux measured using this method is $F_{\rm GeV}$ = 39.1$\pm$2.5, thereby making this peculiar $\gamma$-ray flare as the brightest event recorded from 3C 279 not only since the beginning of {\it Fermi} operation but also ever. For a comparison, the historically brightest flare was detected by {\it EGRET} when the measured flux was as high as $\approx$2 $\times$ 10$^{-5}$ \phflux \citep{1998ApJ...497..178W}. Further, the associated GTI bin size is $\sim$14 min with TS = 2407 and 275 counts are registered.

\begin{figure*}
\hbox{
      \includegraphics[width=\columnwidth]{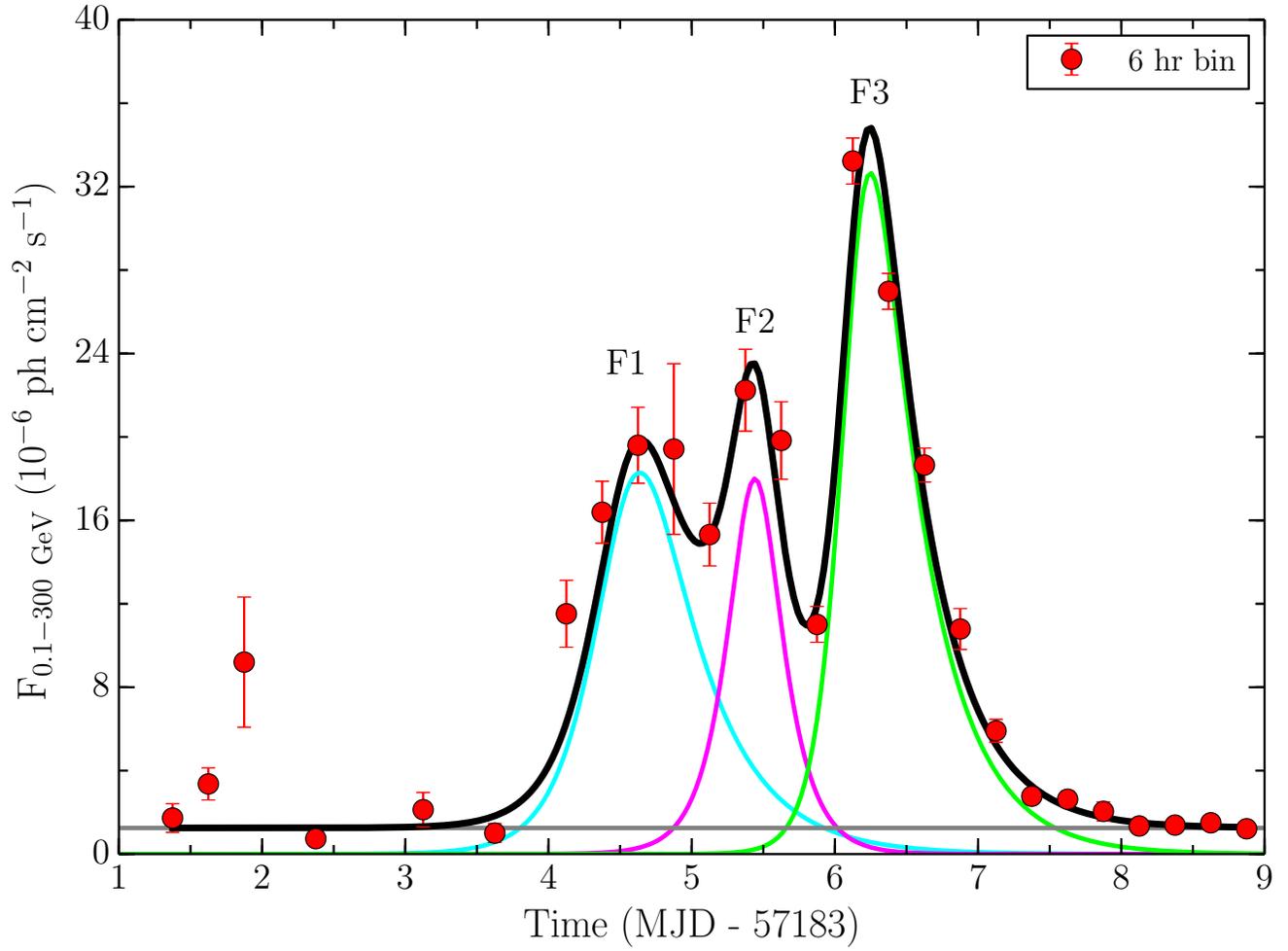}
     }
\caption{Temporal profile fitting of 6 hr binned $\gamma$-ray light curve. Three flares are shown by cyan, pink, and limegreen colors, whereas grey line represnts the constant background. Black line is the sum of all the components.}\label{fig:flare_fitting}
\end{figure*}

To determine the duration of the shortest flux variability, the light curves are scanned using the following equation
\begin{equation}
F(t_2) = F(t_1).2^{(t_2-t_1)/\tau_{\rm d}}
\end{equation}
where $F(t_1)$ and $F(t_2)$ are the fluxes at time $t_1$ and $t_2$ respectively, and $\tau_{\rm d}$ is the flux doubling/halving timescale. The shortest flux doubling time using this method is measured as 2.2$\pm$0.3 hr on MJD 57,189 with $\sim$9$\sigma$ significance. Moreover, in order to assess the asymmetry of rise and decay of flares, all the three flares (F1, F2, and F3) are also subjected to temporal profile fitting. This is done for 6 hr binned flares and the fitting is performed by including a constant background and three fast temporally evolving components. Each flare is approximated to be fit by a function of the following form
\begin{equation}
F(t) = 2F_0 \left[{\rm exp}\left(\frac{t_{\rm 0} - t}{T_{\rm r}}\right) + {\rm exp}\left(\frac{t-t_{\rm 0}}{T_{\rm d}}\right)\right]^{-1}
\end{equation}
\citep{2010ApJ...722..520A}, where $F_0$ is the flux at time $t_0$ representing approximately the flare amplitude, and $T_{\rm r}$ and $T_{\rm d}$
are the rise and decay time of the flare. The obtained rising time for F1, F2, and F3 are 5.40$\pm$0.84 hr, 4.05$\pm$2.73 hrs, and 3.06$\pm$0.57 hr respectively, whereas the associated decay time of the flares are 9.91$\pm$5.04 hr, 4.13$\pm$2.02 hr, and 8.68$\pm$0.42 hr respectively. The result of the time profile fitting is shown in Figure~\ref{fig:flare_fitting}.

\begin{figure*}
\hbox{\hspace*{1cm}
      \includegraphics[width=14cm]{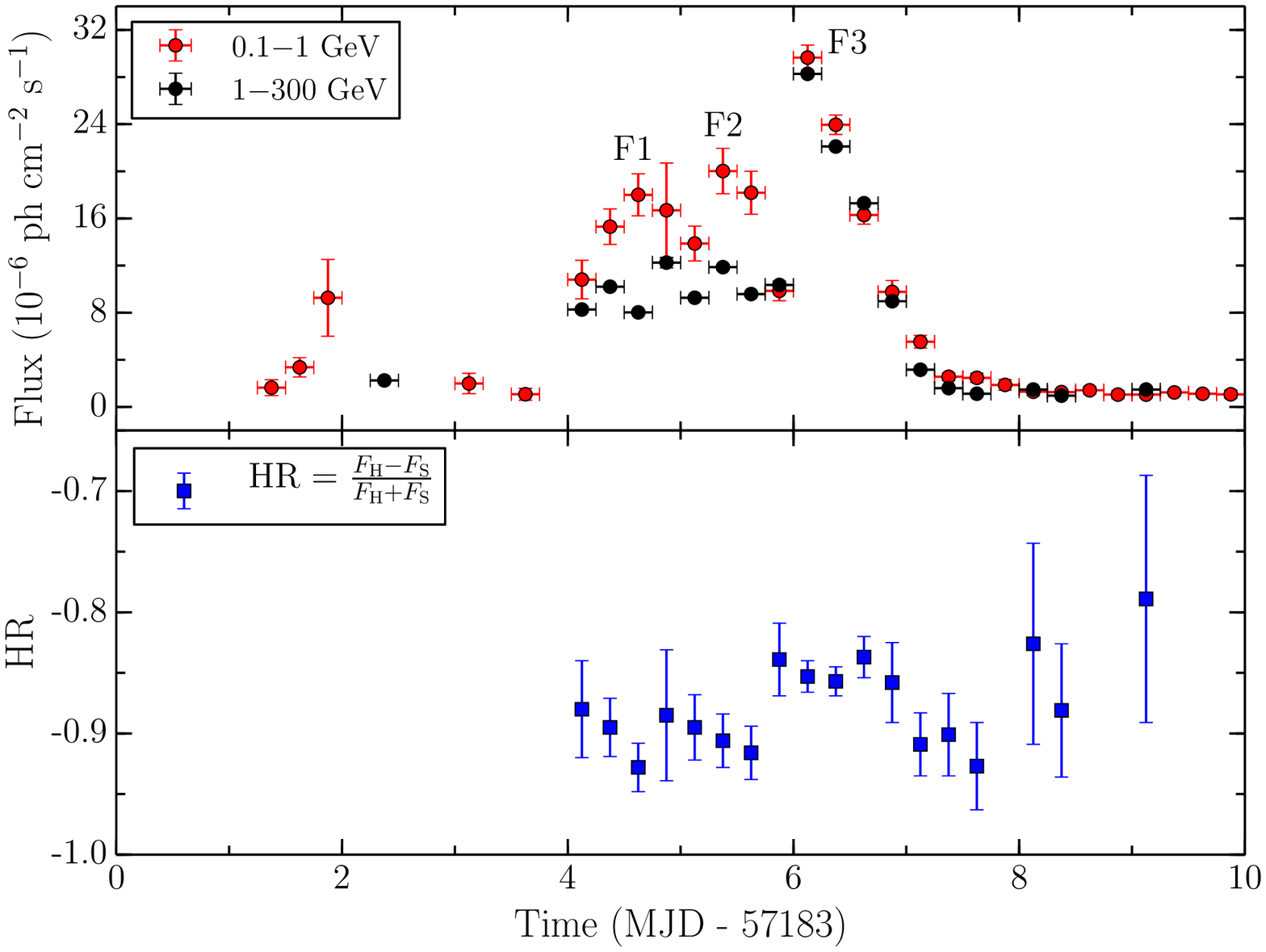}
      }
\hbox{\hspace*{1cm}    
      \includegraphics[width=14cm]{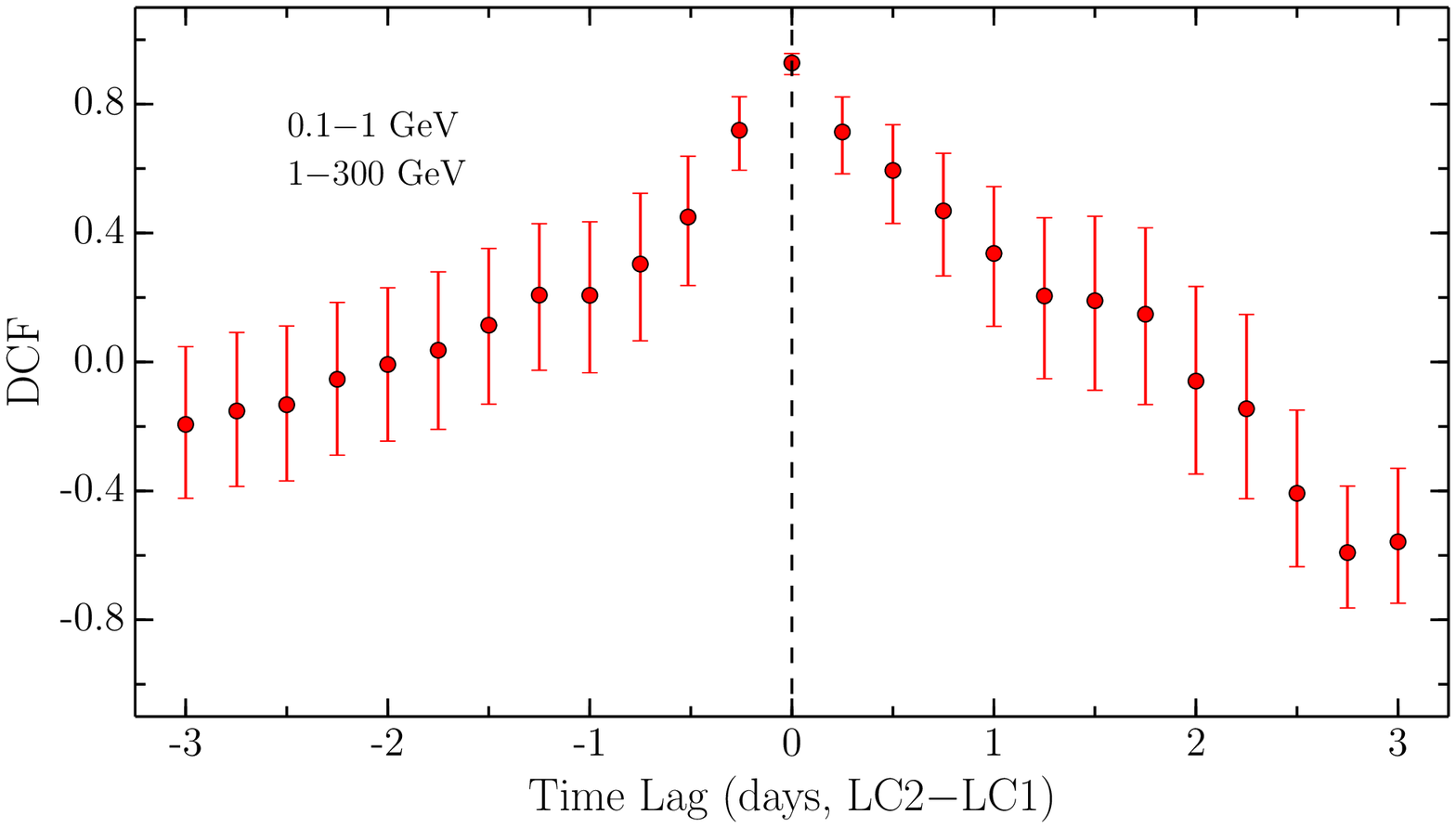}
     }
\caption{Top: Six hr binned light curves of the fluxes in 0.1$-$1 GeV (red) and 1$-$300 GeV, multiplied by a factor of 12 (black). Middle panel presents the variation of HR. Bottom: DCF calculated for 0.1$-$1 GeV and 1$-$300 GeV light curves. The time ordering is T((1$-$300 GeV)$-$(0.1$-$1 GeV)).}\label{fig:HR}
\end{figure*}

LAT $\gamma$-ray light curves of 3C 279, focusing on the flare period, in two different energy bands: 0.1$-$1 GeV and 1$-$300 GeV are shown in the top panel of Figure~\ref{fig:HR}. The 1$-$300 GeV light curve is scaled appropriately to compare its variability pattern with that seen in 0.1$-$1 GeV range. As can be seen, though the pattern of flux variations are similar in both the bands during F3 flare, there are very moderate change in source brightness in 1$-$300 GeV band compared to 0.1$-$1 GeV for F1 and F2 events. This observation indicates that the flare F1 and F2 are primarily caused by low energy electrons. Further, the hardness ratio is also calculated to assess the spectral evolution as a function of time. This is done using following equation
\begin{equation}\label{eq:HR}
{\rm HR} = \frac{F_{\rm H} - F_{\rm S}}{F_{\rm H} + F_{\rm S}},
\end{equation}
where $F_{\rm S}$ and $F_{\rm H}$ are 6 hr binned $\gamma$-ray fluxes in 0.1$-$1 GeV and 1$-$300 GeV energy ranges, respectively. Though there are no significant spectral hardening/softening during F1 and F2 flares, the spectrum is clearly hard at the peak of F3 and it softens as the flare decays (see middle panel of Figure~\ref{fig:HR}). Further, recently it has been proposed that the origin of $\gamma$-ray flares to be lying inside/outside the BLR is reflected from the absence/presence of time lag between MeV and GeV emission \citep{2012ApJ...758L..15D}. To test this hypothesis, a time lag analysis is performed using the {\it z-transformed discrete correlation function} (ZDCF) method of \citet{2013arXiv1302.1508A} \citep[see also][]{1997ASSL..218..163A}. The errors are computed using a Monte Carlo simulation by adding a random error at each step to each data from the errors in the light curves \citep[][]{2013arXiv1302.1508A}. The result of this analysis is shown in bottom panel of Figure~\ref{fig:HR}. As can be seen, there is no significant lead/lag noticed between two energy bands (time lag of 0.0$^{+0.1}_{-0.1}$ days). It should be noted that there could be lag between these light curves, however, limited time resolution of the light curves and closeness of the LAT energy bands makes it difficult to quantify.

The hints for the presence of spectral evolution can be further investigated by plotting photon index versus flux. This is done for 6 hr binned data, in two energy ranges: 0.1$-$300 GeV and 0.1$-$1 GeV, for the period P1 and P2 (covering the flares F1, F2, and F3). The results are presented in the left panel of Figure~\ref{fig:gamma_spec}. For comparison, the averages obtained over the period `Pre', is also shown with blue squares. A clockwise pattern is evident during P2 period for both energy bands. Though there are hints for the same behavior during P1 period in 0.1$-$300 GeV energy range, the 0.1$-$1 GeV plot show a clear evidence of clockwise helical pattern. The reduced $\chi^2$ for a constant fit of the photon index for the period P1 are 6.94/7 and 6.68/7, for 0.1$-$300 GeV and 0.1$-$1 GeV respectively. Comparing to `Pre' period, a flux enhancement by a factor of 10 is accompanied by slight hardening of the spectrum. While drawing any firm conclusion on particle acceleration and cooling from these patterns is not possible, the lack of strong spectral variability still provides clues to the underlying radiative processes. Due to poor statistics, a clear pattern of the photon index for energy range 1$-$300 GeV versus flux could not be observed with 6 hr time binning.

In the right panel of Figure~\ref{fig:gamma_spec}, the LAT spectra obtained by averaging the four periods shown in Figure~\ref{fig:fermi_lc}, are displayed. These flux distributions have been fitted with PL (dotted), log-parabola (dashed), and BPL (solid) models. Due to low photon statistics during the periods `Pre' and `Post', the BPL model could not be fitted. To substantiate the presence/absence of the spectral curvature, the TS of the curvature $TS_{\rm curve}$ = 2(log $\mathcal{L}$(log-parabola/BPL) $-$ log $\mathcal{L}$(PL)) is also computed \citep{2012ApJS..199...31N}. The associated parameters are provided in Table~\ref{tab:gamma_spec}. Out of all the four periods considered, a statistical significant break/curvature is noticed only during P1 and P2 periods. Both the log-parabola and BPL model reproduce the data satisfactorily, though the log-parabola model gives slightly better fit. The break energy obtained from the BPL model fitting remains constant irrespective of the fluctuations seen in the flux levels, during both the periods. This finding is inline with that observed during the $\gamma$-ray outburst of 3C 454.3 in 2010 November \citep{2011ApJ...733L..26A}.

\section{Discussion}\label{sec:dscsn}
The $\gamma$-ray outburst of 3C 279 in 2015 June has broken the record of the highest $\gamma$-ray flux ever measured from this blazar. The entire flaring episode is found to comprise of three flares with the last one having the largest amplitude. The rising part of the brightest flare is unresolved down to 3 hr binning and this hints that with the limited time resolution, the apparent profiles of the source flux variations may not reflect the exact temporal characteristics of the $\gamma$-ray outburst. However, the decaying part of the flare is nicely resolved in 6 hr and 3 hr binned light curves.

The shortest flux doubling timescale obtained in this work is $\sim$2 hr which is similar to that obtained during 2013 December outburst of 3C 279 \citep{2015arXiv150204699H}. Interestingly, comparing to 2014 April outburst, though the flux level is significantly higher during 2015 June flare, the faster variability was measured during the former \citep{2015ApJ...803...15P}. However, similar to 2013 December and 2014 April flares, the flares of 2015 June outburst are found to exhibit asymmetric variability patterns with fast rise and slow decay (except for F2), which can be explained by invoking the rapid injection of accelerated electrons, probably at shock front, and the decay can be attributed to the weakening of the shock.

\begin{figure*}
\hbox{
      \includegraphics[width=15cm]{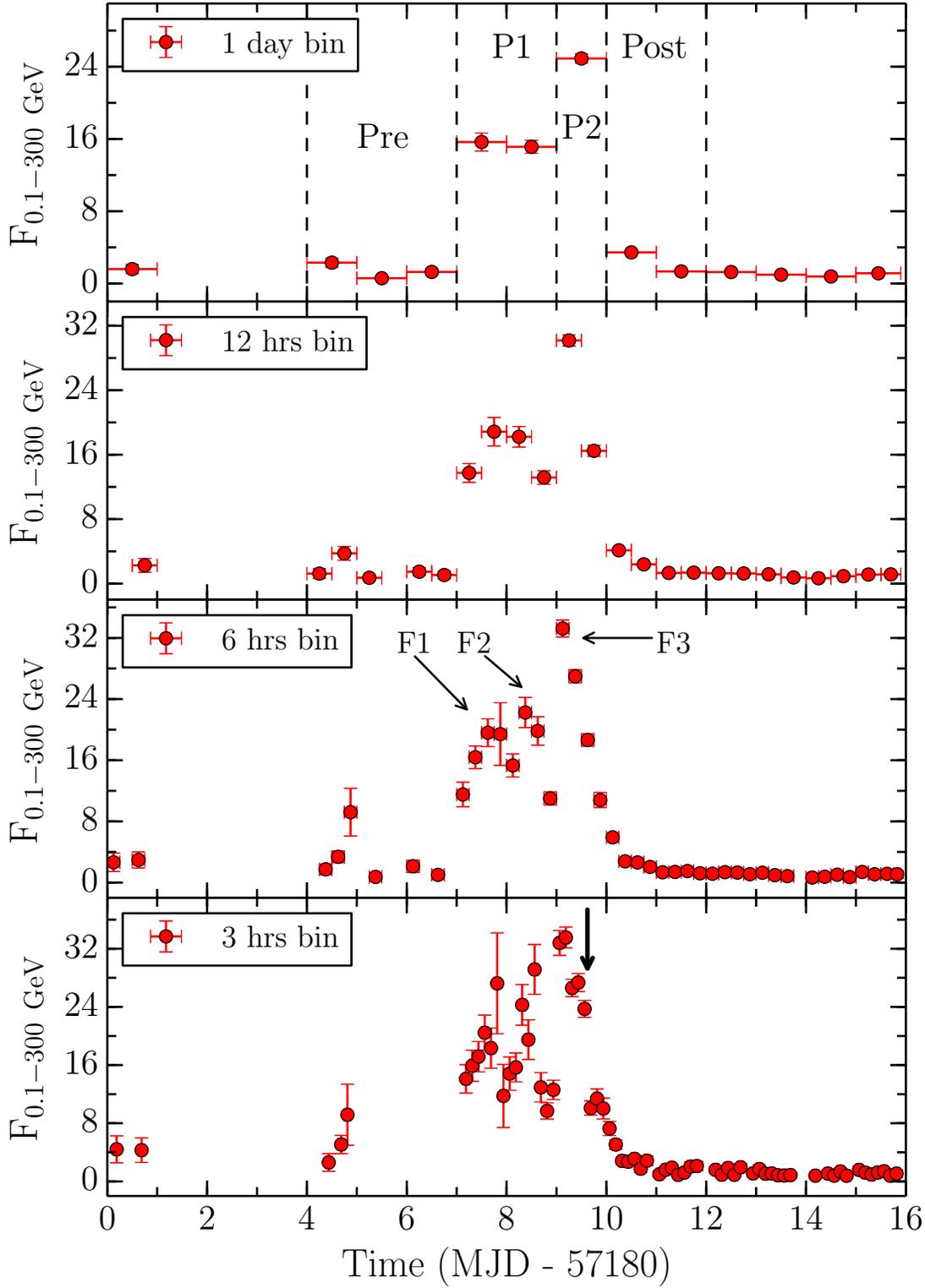}
     }
\caption{Gamma-ray light curve of 3C 279 covering the period of outburst. Fluxes are in units of 10$^{-6}$ \phflux. See text for details.}\label{fig:fermi_lc}
\end{figure*}

\begin{figure*}
\hbox{\hspace*{-1cm}
      \includegraphics[width=10cm]{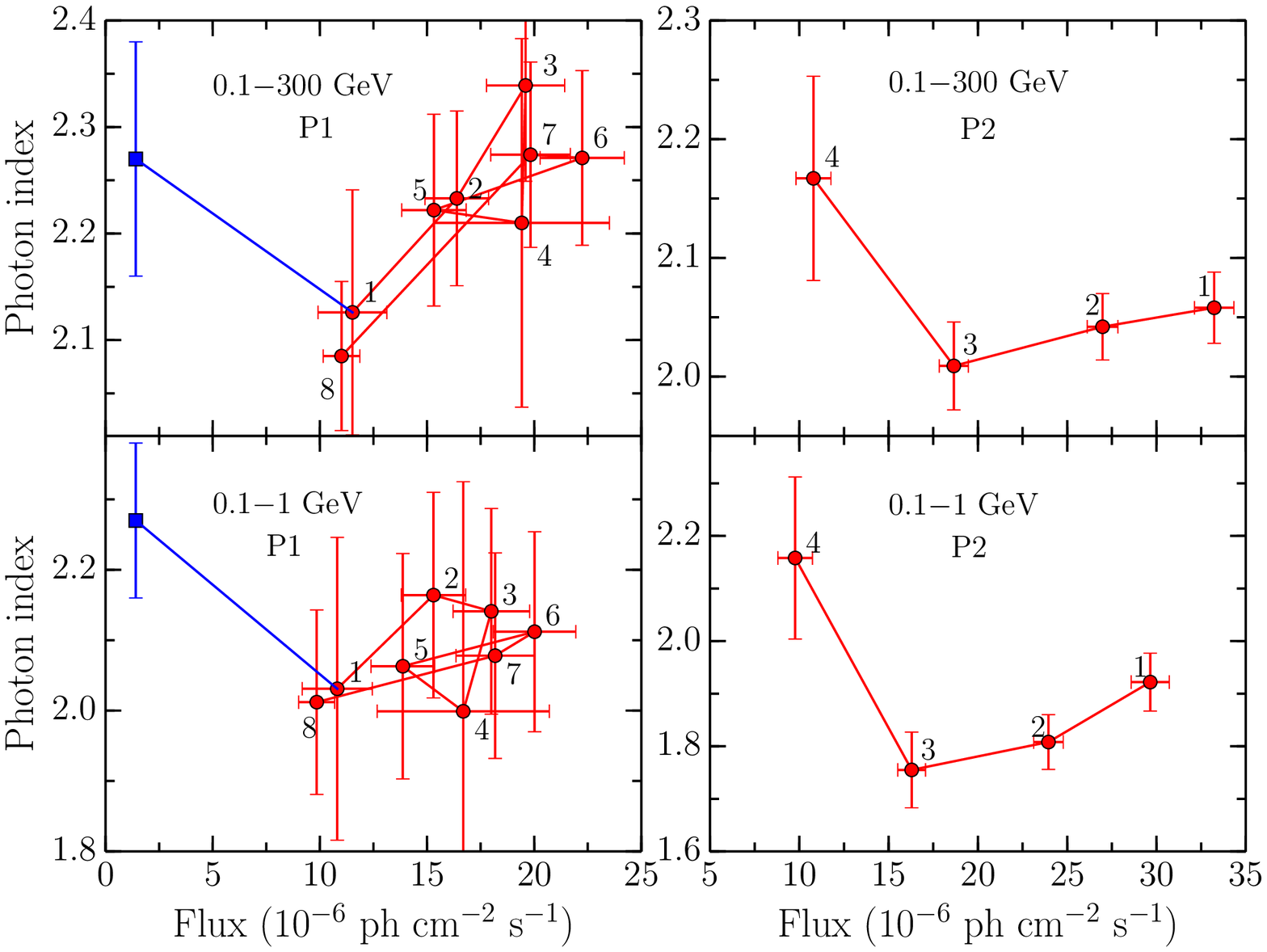}
      \includegraphics[width=10cm]{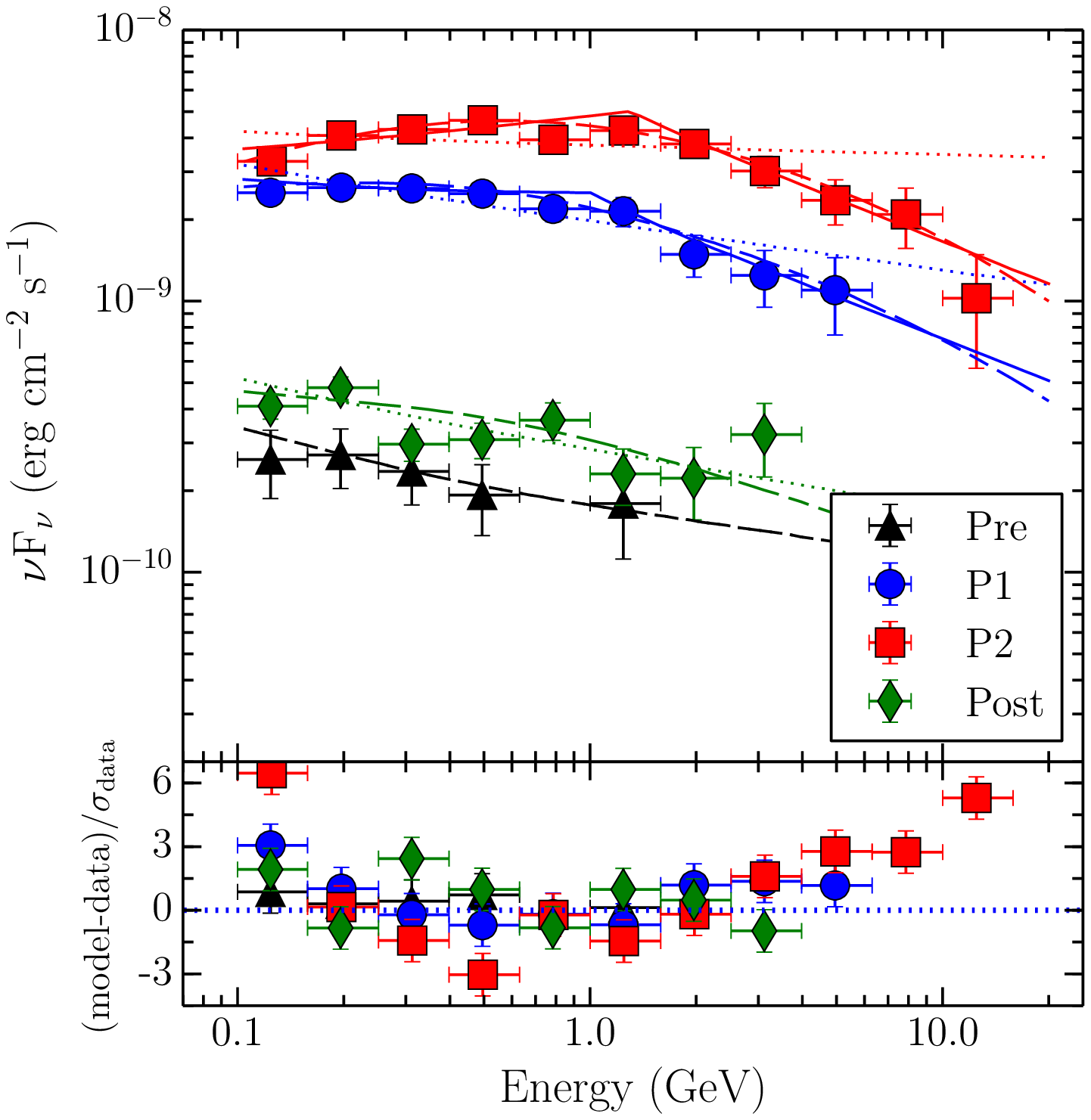}
     }
\caption{Left: $\Gamma$ vs. flux for the periods P1 and P2, obtained with 6 hr binning for 0.1$-$300 GeV and 0.1$-$1 GeV bands. Right: {\it Fermi}-LAT SEDs during different activity states as defined in Figure~\ref{fig:fermi_lc}. PL, LP, and BPL models are shown with dotted, dashed, and solid lines respectively. Horizontal errorbar corresponds to energy range of each bin whereas vertical bar represents 1$\sigma$ statistical errors. The residuals in the lower panel refer to the PL model.}\label{fig:gamma_spec}
\end{figure*}

The minimum Doppler factor $\delta_{\rm min}$ can be estimated numerically from $\gamma\gamma$ opacity arguments and by measuring the energy of the highest energy photon detected during the flare. Assuming that the optical depth $\tau_{\gamma\gamma}$($\epsilon_1$) of a photon with energy $\epsilon_1 = E_1/m_ec^2$ to the $\gamma\gamma$ interaction is $\tau_{\gamma\gamma}$ = 1, the minimum Doppler factor can be calculated as follows \citep[see e.g.][]{1995MNRAS.273..583D,2010ApJ...716.1178A}
\begin{equation}
\delta_{min}\cong \left[\frac{\sigma_T d_L^2(1 + z)^2 f_{\hat{\epsilon}} \epsilon_1}{4t_{var}m_ec^4}\right]^{1/6},
\end{equation}
 where $f_{\epsilon}$ is the flux at energy $\epsilon = h\nu/m_ec^2$ and $\epsilon=\hat{\epsilon}=2\delta^2/(1+z)^2\epsilon_1$ \citep{2010ApJ...716.1178A}. In this calculation, the energy of the highest energy photon ($E_1$) is used during the period when variability time $t_{\rm var}$ and $f_{\epsilon}$ are measured. The highest energy photon of energy $\approx$52 GeV is detected on MJD 57,189.62 at an angular separation of 0$^{\circ}$.09 from the 3FGL position of 3C 279, with 99.99\% probability of being source photon (see bold downward arrow in the bottom panel of Figure~\ref{fig:fermi_lc}). Almost at the same time, the {\it Swift} X-ray Telescope observation has revealed the energy flux in 0.3$-$10 keV band as 3.5 $\times$ 10$^{-11}$ \ergflux~and the associated photon index is 1.4 \citep{2015ATel.7668....1P}. Since the highest energy photon was detected during the fading part of F3, the variability time can be taken as $t_{\rm var}=$ ln(2) $\times T_d\approx6$ hr. This gives $\delta_{\rm min}\approx$ 14.

A rough estimation about the location of the emission region ($R_{\rm diss}$) can be done by assuming it to have spherical geometry and bulk Lorentz factor $\Gamma=\delta_{\rm min}=14$. Then $R_{\rm diss}<2c\Gamma^2t_{\rm var}/(1+z)\approx$0.05 pc. Now, considering the accretion disk luminosity ($ L_{\rm disk}$) of 3C 279 as 1 $\times$ 10$^{45}$ \lum~\citep{2015ApJ...803...15P}, the size of the BLR comes out to be $\approx$1 $\times$ 10$^{17}$ cm $\approx$0.03 pc \citep[e.g.][]{2007ApJ...659..997K}. Therefore, at the time of the 52 GeV photon emission, the $\gamma$-ray radiating region must have been located either close to the outer edge of the BLR or beyond it. This is also supported by fact that to avoid severe absorption of high energy photons (having energy $>$50  GeV) by optical-UV radiation of BLR radiation field, the emitting region has to be farther out from the BLR.

The highest $\gamma$-ray flux measured is $F_{\rm GeV}$ = 39.1$\pm$2.5 and the associated photon index is 2.0$\pm$0.1. This corresponds to an isotropic luminosity ($L_{\gamma}$) of (5.5$\pm$1.1) $\times$ 10$^{49}$ \lum. Assuming $\Gamma=14$, the $\gamma$-ray luminosity in the jet frame would be $L_{\gamma, em} \simeq L_{\gamma}/2\Gamma^{2} \simeq$ 1.4 $\times$ 10$^{47}$ \lum. This is about 10 times the total available accretion power ($L_{\rm acc} \simeq L_{\rm disk}/\eta_{\rm disk} \simeq 1 \times 10^{46}$ erg s$^{-1}$; assuming radiative efficiency $\eta_{\rm disk}$ = 10\%). The black hole mass of 3C 279 is found in the range of (3$-$8) $\times$ 10$^{8}M_{\odot}$ \citep{2001MNRAS.327.1111G,2002ApJ...579..530W}, and thus the Eddington luminosity $L_{\rm Edd}\approx$ (0.4$-$1) $\times$ 10$^{47}$ \lum. Now, in order for the observed $L_{\gamma}$ to be lower than $L_{\rm Edd}$, $\theta_j\lesssim5^{\circ}-7^{\circ}$ and a beaming factor (1 $-$ cos $\theta_j$)$^{-1}\gtrsim240-440$. Similar parameters are reported by \citet{2011ApJ...733L..26A} for the GeV flare of 3C 454.3 during its 2010 November flare.

The cooling timescale of the $\gamma$-ray ($\epsilon_{\gamma}=1$ GeV) emitting electrons can be calculated as follows \citep[e.g.][]{2013ApJ...766L..11S}
\begin{equation}
\tau_{\rm cool} \simeq \frac{3m_{e}c}{4\sigma_{\rm T}u'_{\rm BLR/torus}} \sqrt{\frac{\epsilon_0(1+z)}{\epsilon_{\gamma}}},
\end{equation}
where $u'_{\rm BLR/torus}=f_{\rm BLR/torus}L_{\rm disk}\Gamma^2/4\pi R^2_{\rm BLR/torus}$, is the comoving seed photon energy density for the BLR/torus radiation field with characteristic energy $\epsilon_0=$ 10.2/0.2 eV and $f_{\rm BLR/torus}$ is the fraction of $L_{\rm disk}$ reprocessed by BLR/torus. This resulted in $\tau_{\rm cool}=$ 7 min or 11 min, provided the seed photons for EC process are originated from the BLR or torus, respectively. The obtained cooling time is significantly shorter than that observed decay timescale of the flares, suggesting that the observed flare decrease is controlled not only by radiative cooling, but by a combination of various factors such as the geometry and sub-structure of the emission region or possibly the jet dynamics \citep{2001ApJ...563..569T,2014MNRAS.442..131K}.

A significant break/curvature in the $\gamma$-ray spectrum is the characteristic properties of powerful FSRQs. Interestingly, this feature is found to be more prominent during the flaring episodes \citep[see e.g.][]{2011ApJ...733L..26A,2015ApJ...803...15P}. During P2 period, i.e. at the peak of the flare, the BPL photon index $\Gamma_1$ before $E_{\rm break}$ ($\approx$1 GeV) is $<2$, thus indicating a rising spectrum, whereas $\Gamma_2>2$, implying a falling spectrum after $E_{\rm break}$. This observation suggests that probably the peak of the IC mechanism is seen in the LAT bandpass and thus the observed shape most likely reflects the energy distribution of the emitting electrons. Comparing the shape of the $\gamma$-ray spectrum with that seen during 2013 December and 2014 April reveals an interesting observation. The 2013 December flare exhibited a hard rising spectrum \citep[$\Gamma=1.71\pm0.10$,][]{2015arXiv150204699H} indicating that the IC mechanism to peak at very high frequencies, whereas 2014 April events showed a steep falling spectrum \citep[{\bf $\Gamma=2.23\pm0.03$},][]{2015ApJ...803...15P}, thus implying IC peak to lie before the LAT energy range. These three flares, therefore, represent the three different characteristics of the radiative processes powering the relativistic jet of 3C 279. The observation from {\it Fermi}-LAT, in this regard, is an invaluable asset to learn the physical properties of the most powerful objects in the Universe, called blazars.

\acknowledgments

\begin{table}
\caption{Results of the Model Fitting to the $\gamma$-ray Spectra of 3C 279, obtained for different activity periods.}\label{tab:gamma_spec}
\begin{center}
\begin{tabular}{lcccccc}
\hline
         &                         & Power law                    &    &    &    &    \\
Activity & $F_{\rm GeV}$ & $\Gamma_{0.1-300~{\rm GeV}}$ &    &    & TS &              \\
         & (10$^{-6}$\phflux)      &                              &    &    &    &    \\ 
\hline
Pre      & 1.41$\pm$0.19  & 2.27$\pm$0.11 & --          & --                  & 274.55   & --   \\
P1       & 15.80$\pm$0.55 & 2.21$\pm$0.03 & --          & --                  & 6553.42  & --   \\
P2       & 24.50$\pm$0.49 & 2.05$\pm$0.02 & --          & --                  & 22673.92 & --   \\
Post     & 2.39$\pm$0.13  & 2.26$\pm$0.05 & --          & --                  & 1690.61  & --   \\
\hline
         &                    & Log parabola &    &        &    &                           \\
Activity & $F_{\rm GeV}$      & $\alpha$     & $\beta$     &    & TS      & $TS_{\rm curve}$ \\
         & (10$^{-6}$\phflux) &              &             &    &         &                  \\ 
\hline
Pre      & 1.41$\pm$0.19        & 2.27$\pm$0.11  & 0.00$\pm$0.00 & -- & 274.55   & 0.00       \\
P1       & 15.20$\pm$0.55       & 2.05$\pm$0.05  & 0.11$\pm$0.03 & -- & 6572.25  & 18.83     \\
P2       & 23.60$\pm$0.48       & 1.84$\pm$0.03  & 0.13$\pm$0.02 & -- & 22769.93 & 96.01     \\
Post     & 2.34$\pm$0.13        & 2.16$\pm$0.08  & 0.08$\pm$0.04 & -- & 1694.40  & 3.79      \\
\hline
         &                    & Broken power law &    &        &    &                           \\
Activity & $F_{\rm GeV}$      & $\Gamma_1$     & $\Gamma_2$    & $E_{\rm break}$     & TS      & $TS_{\rm curve}$ \\
         & (10$^{-6}$\phflux) &                &               & (GeV)               &         &                  \\ 
\hline
P1       & 15.31$\pm$0.55       & 2.05$\pm$0.05    & 2.60$\pm$0.11   & 1.00$^{+0.35}_{-0.30}$ & 6571.30  & 17.88 \\
P2       & 23.85$\pm$0.48       & 1.88$\pm$0.03    & 2.58$\pm$0.07   & 1.30$^{+0.10}_{-0.15}$ & 22757.23 & 83.31 \\
\hline
\end{tabular}
\end{center}
\end{table}

\newpage

\newpage

\newpage

\newpage

\end{document}